\documentclass[prd,preprint,showpacs,preprintnumbers,amsmath,amssymb]{revtex4}
\usepackage{mathrsfs}
\DeclareMathAlphabet{\mathpzc}{OT1}{pzc}{m}{it}

\usepackage[english]{babel}

\usepackage{amsmath}
\usepackage{amssymb}

\usepackage{mathrsfs}
\DeclareMathAlphabet{\mathpzc}{OT1}{pzc}{m}{it}

\usepackage{graphicx}

\usepackage{cancel} % Cro
\usepackage{paralist}
\usepackage{booktabs}
\usepackage{hyperref}
\usepackage{xspace}

\usepackage{graphicx,color,verbatim,acronym,epsfig,subfigure}
\usepackage{slashed,skak}
\usepackage{latexsym,amsfonts,amssymb,amsmath,makeidx,mathrsfs,multirow,lineno,bm,bbm}

\newcommand{\abs}[1]{|#1|}

\begin{document}
\title{Gravitational Waves from first-order phase transition and domain wall}

\author{Ruiyu Zhou }

\author{Jing Yang}
\author{Ligong Bian }
\email{lgbycl@cqu.edu.cn}
\affiliation{
	~Department of Physics, Chongqing University, Chongqing 401331, China}
\date{\today}

\begin{abstract}

 In many particle physics models, domain wall can form during the phase transition process after discrete symmetry breaking. We study the scenario within a complex singlet extended Standard Model framework, where a strongly first order phase transition can occur depending on the hidden scalar mass and the mixing between the extra heavy Higgs and the SM Higgs mass. The gravitational wave spectrum is of a typical two-peak shape, the amplitude and the peak from the strongly first order phase transition is
able to be probed by the future space-based interferometers, and the one locates around the peak from the domain wall decay is far beyond the capability of the current PTA, and future SKA.

\end{abstract}

\graphicspath{{figure/}}

\maketitle
\baselineskip=16pt

\pagenumbering{arabic}

\vspace{1.0cm}
\tableofcontents

\newpage

\section{Introduction}

The observation of black hole binary merger~\cite{Abbott:2016blz} and the approval of Laser Interferometer Space Antenna (LISA) by European Space Agency~\cite{Audley:2017drz} raise growing interest in the gravitational wave study. The strongly first order Electroweak phase transition, as one of the crucial ingredients for the Electroweak baryogenesis~\cite{Morrissey:2012db}, can provide a detectable stochastic gravitational wave background with the peak frequency within the sensitivity of the LISA~\cite{Caprini:2019egz}. The phase transition in the Standard model with the observed Higgs mass is confirm to be {\it cross-over}~\cite{DOnofrio:2014rug}, and a first-order Electroweak phase transition usually requires extension of the Standard Model Higgs sectors~\cite{Mazumdar:2018dfl}. The Higgs pair searches at future collider can serve as a probe of the phase transition parameters of new physics models~\cite{Arkani-Hamed:2015vfh}. That make it possible to search the strongly first order phase transition in particle physics models with collider and and gravitational wave complementary~\cite{Chen:2019ebq,Alves:2019igs,Alves:2018jsw,Hashino:2018wee,Hashino:2016rvx,Bian:2019zpn}.
The cosmic phase transition with spontaneously broken of a discrete symmetry may create domain walls~\cite{Kibble:1976sj}, which can be unstable to avoid overclose the Universe~\cite{Vilenkin:1981zs,Gelmini:1988sf,Larsson:1996sp} by including approximate and explicitly broken terms in the models. Different from literatures, we are going to study domain walls formation and decay after a strongly first-order Electroweak phase transition, and study the gravitational waves produced during the process.
Concretely, we study the phase transition with a complex singlet scalar extended Standard model with a $\mathbb{Z}_3$ symmetry. The gravitational wave from the strongly first-order phase transition (with the model) has been study previously in Refs.~\cite{Kang:2017mkl,Kannike:2019mzk,Chiang:2019oms}.
Different from these studies, we study the gravitational waves produced from the strongly first-order Electroweak phase transition, and domain wall decay. We first check the strongly first-order Electroweak phase transition condition by evaluating the baryon number preservation criterion (BNPC), and then study the relation among the criterion and gravitational wave parameters, i.e, the latent heat, and the inverse duration of the phase transition. After that, we study the possibility to have a detectable gravitational wave from the domain wall decay at the European Pulsar Timing Array (EPTA \cite{Desvignes:2016yex}), the Parkes Pulsar Timing Array
(PPTA \cite{Hobbs:2013aka}), and the International Pulsar Timing Array (IPTA \cite{Verbiest:2016vem}). Finally, we observe a gravitational wave signal with one peak locates at these Pulsar Timing Arrays sensitivity range and another peak can be covered by future space-based interferometers.

%%%%%%%%%%%%%%%%%%%%%%%%%%%%%%%%%%%%%%%%%%%%%%%%%%%
\section{The $\mathbb{Z}_{3}$ symmetric complex singlet extended Standard Model}
\label{sec:model}

In this work, we consider the model with the scalar potential being given by,
\begin{align}
  V = \mu_{H}^{2} \abs{H}^{2} + \lambda_{H} \abs{H}^{4}
  + \mu_{S}^{2} \abs{S}^{2} + \lambda_{S} \abs{S}^{4} + \lambda_{SH} \abs{S}^{2} \abs{H}^{2} + \frac{\mu_3}{2} (S^{3} + S^{\dagger 3})\;.
\end{align}
The cubic $\mu_{3}$ term breaks the global $U(1)$ $S \to e^{i \alpha} S$ symmetry with a remanent unbroken $\mathbb{Z}_{3}$ symmetry.
We expand the scalar fields around their classical backgrounds as,
\begin{equation}
  H =
  \begin{pmatrix}
    G^+
    \\
    \frac{ h + i G^0}{\sqrt{2}}
  \end{pmatrix}\;,
  \qquad
  S = \frac{ s + i \chi}{\sqrt{2}}\;,
\end{equation}
and obtain the tree-level potential,
\begin{align}\label{eq:z3singlet}
V_0(h,s,\chi)&=\frac{\lambda_{H}}{4}h^{4}+\frac{\lambda_{S}}{4}s^{4}+\frac{\lambda_S}{4}\chi^{4}
+\frac{\lambda_{SH}}{4} h^{2}s^{2}+\frac{ \lambda_{SH}}{4} h^{2}\chi^{2}+\frac{\lambda_S}{2} s^{2} \chi^{2} \nonumber\\
&+\frac{\mu_3}{2 \sqrt{2}}s^{3}-\frac{3 \mu_3}{2 \sqrt{2}} s\chi^{2}+\frac{\mu_H^{2}}{2}h^{2}+\frac{\mu_s^{2}}{2}s^{2}+\frac{\mu_s^{2} }{2}\chi^{2}\;.
\end{align}
Considering the stationary point conditions,
\begin{align}
\left.\frac{d V_{0}(h, s, \chi)}{d h}\right|_{h=v}=0\;,\left.\frac{d V_{0}(h, s, \chi)}{d s}\right|_{s=v_s}=0\;,
\end{align}
we get $\mu_{H}^2=-\lambda_{H} v^2 - \frac{1}{2} \lambda_{SH} v_s^2$, $\mu_s^2=-\lambda_S v_s^2 - \frac{1}{2} \lambda_{SH} v^2 - \frac{3\sqrt{2}}{4} \mu_3 v_s $.
The Higgs mass matrix is then given by,
\begin{equation}
  M^{2} =
  \begin{pmatrix}
    2 \lambda_{H} v^{2} & \lambda_{SH} v v_{s}
    \\
    \lambda_{SH} v v_{s} & 2 \lambda_{S} v_{s}^{2} + \frac{3}{2 \sqrt{2}} \mu_{3} v_{s}
  \end{pmatrix}.
\label{eq:mass:matrix}
\end{equation}
The stationary point can be a minimum when one have a positive determination of the zero temperature Hessian matrix, which explicitly gives
\begin{equation}
\lambda_H > 0 \;,
8 v_s \lambda_S + 3\sqrt{2} \mu_3 >0\;,
8 v_s \lambda_H \lambda_S -2 v_s \lambda_{SH}^{2}+3 \sqrt{2} \lambda_H \mu_3>0\;.
\end{equation}
Introducing the rotation matrix $R=((\cos \theta, \sin \theta),(-\sin \theta, \cos \theta))$, and rotating into the mass basis through
\begin{equation}
\left(\begin{array}{l}{h_{1}} \\ {h_{2}}\end{array}\right)=\left(\begin{array}{cc}{\cos \theta} & {-\sin \theta} \\ {\sin \theta} & {\cos \theta}\end{array}\right)\left(\begin{array}{l}{h} \\ {s}\end{array}\right)\;,
\end{equation}
 one has,
\begin{align}
&m_{h_1}^2 = \frac{1}{4} v_s (8 v_s \lambda_S+3 \sqrt{2} \mu_3) \sin\theta^{2} +2 v \cos\theta(v \lambda_H \cos\theta+v_s \lambda_{SH} \sin\theta)\;, \nonumber\\
&m_{h_2}^2 = \frac{1}{4} v_s (8 v_s \lambda_S+3 \sqrt{2} \mu_3) \cos\theta^{2} - 2 v v_s \lambda_{SH} \cos\theta \sin\theta+2 v^{2} \lambda_H \sin\theta^{2}\;.
\end{align}
The mixing angle $\theta$ can be expressed as follows,
\begin{equation}
   \tan 2 \theta = \frac{\lambda_{SH}  v v_{s}}{\lambda_{H} v^{2} - \lambda_{S} v_{s}^{2} - \frac{3}{4 \sqrt{2}} \mu_{3} v_{s}}.
\end{equation}
For our study, we consider $h_1=h_{SM}$, and $m_{h_2}>m_{h_1}$.
After Electroweak symmetry together with the $\mathbb{Z}_{3}$ symmetry breaking, the $S \to S^{\dagger}$ is equivalent to $\chi \to -\chi$. Different from Ref.~\cite{Chiang:2019oms,Kannike:2019mzk}, we do not consider $\chi$ as dark matter in this work.
The mass of the pseudo-Goldstone $\chi$ is given by,
\begin{equation}
  m_{\chi}^{2} = - \frac{9}{2 \sqrt{2}} \mu_{3} v_{s},
\end{equation}
which is proportional to $\mu_{3}$ as it explicitly breaks the $U(1)$ symmetry.
Requiring the Electroweak together with $\mathbb{Z}_{3}$ broken vacuum being the global minimum, one has
\begin{equation}\label{eq:bound}
  m_{\chi}^{2} < \frac{9 m_{h_1}^{2} m_{h_2}^{2}}{m_{h_1}^{2} \cos^2 \theta + m_{h_2}^{2} \sin^{2} \theta}\;,
\end{equation}
which severely constrain the relation among $m_{h_2}, m_{\chi}$, and $\theta$.

The relation between the interaction coupling and physical parameters of Higgs masses, VEVs,  and mixing angle $\theta$ are given as,
\begin{align}
  \lambda_{H} &= \frac{ m_{1}^{2} + m_{2}^{2} + (m_{1}^{2} - m_{2}^{2}) \cos 2 \theta }{4 v^{2}}\;,
  \\
  \lambda_{S} &= \frac{3(m_{1}^{2} + m_{2}^{2}) + 2 m_{\chi}^{2} + 3 (m_{2}^{2} - m_{1}^{2}) \cos 2 \theta}{12 v_{s}^{2}}\;,
  \\
  \lambda_{SH} &= \frac{(m_{1}^{2} - m_{2}^{2}) \sin 2 \theta}{2 v_{s} v}\;,
  \\
  \mu_{H}^{2} &= -\frac{1}{4} (m_{1}^{2} + m_{2}^{2}) + \frac{1}{4 v} (m_{2}^{2} - m_{1}^{2})  (v \cos 2 \theta + v_{s} \sin 2 \theta)\;,
  \\
  \mu_{S}^{2} &= -\frac{1}{4} (m_{1}^{2} + m_{2}^{2}) + \frac{1}{6} m_{\chi}^{2} + \frac{1}{4 v_{s}}
  (m_{1}^{2} - m_{2}^{2})(v_{s} \cos 2 \theta - v \sin 2 \theta)\;,
  \\
  \mu_{3} &= -\frac{2 \sqrt{2}}{9} \frac{m_{\chi}^{2}}{v_{s}}\;.
\end{align}

%The relative depth of the true minimum ($E_\alpha^4$) and  the height of the barrier that separates it from the false one ($E_m^4$) are two characteristics of the phase transition. Both of those quantities are extreme points of the effective temperature potential. With this condition $\partial V_{eff}[T]/\partial h =0$, $E_\alpha^4$ and $E_m^4$ can be numerical calculated

%%%%%%%%%%%%%%%%%%%%%%%%%%%%%%%%%%%%%%%%%%%%%%%%%%%

\section{Electroweak phase transition}
In this section, we first investigate the phase transition dynamics relevant for the domain wall formation. After that, we evaluate the strongly first-order Electroweak phase transition condition given by BNPC.

\subsection{phase transition dynamics}

Utilizing the gauge invariant approach~\cite{Zhou:2018zli,Bian:2019zpn,Bian:2019kmg,Bian:2018bxr,Alves:2018jsw,Bian:2018mkl,Chao:2017vrq}, the finite temperature potential adopted for the study of phase transition behavior in the $\mathbb{Z}_3$ Complex Singlet model is given by
\begin{align}
V_T=\frac{(\mu_H^{2}+c_{hT}) h^{2} }{2}+\frac{(\mu_s^{2} +c_{sT})s^{2} }{2}+\frac{\mu_3 s^{3} }{2 \sqrt{2}}+\frac{\lambda_H h^{4} }{4}+\frac{\lambda_S s^{4}}{4}+\frac{\lambda_{SH} h^{2} s^{2} }{4}\;,
\end{align}
with the finite temperature corrections are calculated as
\begin{align}
c_{hT}&=\frac{1}{48} T^{2}\left(9 g^{2}+3 g'^{2}+4\left(3y_t^{2}+6 \lambda_{H}+\lambda_{SH}\right)\right)\;,\\ c_{sT}&=\frac{1}{6} T^{2}(2\lambda_{S}+\lambda_{S H})\;.
\end{align}

\begin{figure}[!htp]
\begin{center}
\includegraphics[width=0.4\textwidth]{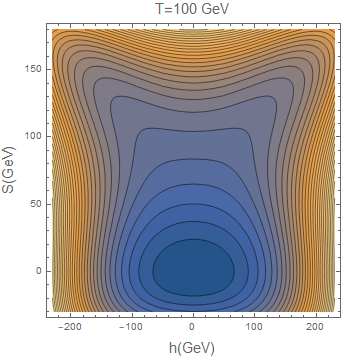}
\includegraphics[width=0.4\textwidth]{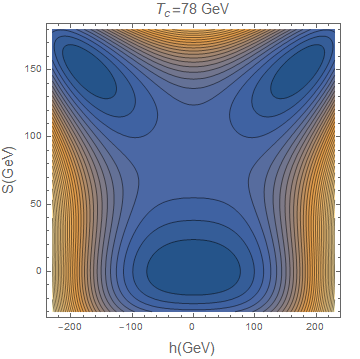}
\includegraphics[width=0.4\textwidth]{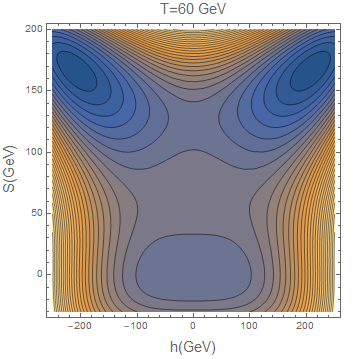}
\includegraphics[width=0.4\textwidth]{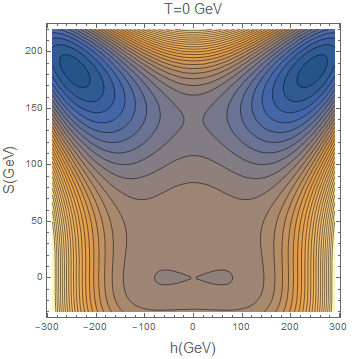}
\caption{This is the phase transition process of $BM1$ (given in Table~\ref{tab:BP}).  }\label{onestep}
\end{center}
\end{figure}

%\subsubsection{one step phase transition}
Depending on the vacuum structure at the zero temperature, there are two different phase transitions types, which are one-step PT $(0,0) \to (h,s)$ and two-step PT $(0,0) \to (h,0)/(0,s) \to (h,s)$, see Appendix.~\ref{sec:apa} for details.
In this study, we focus on the one-step phase transition type. When the temperature of the Universe drops to the critical temperature, one have two potential degenerate at the vacua $O:(0,0)$ and $B:(h,s)$ occurring with a potential barrier structure. Where, one have
\begin{align}
&V_{T}(0,0, T_{C})=V_{T}(h_{B},s_{B}, T_{C})\;,
\frac{dV_{T}(h,s, T_{C})}{dh}|_{h=h_{B},s=s_{B}} = 0 \;,\nonumber\\
&\frac{dV_{T}(h,s, T_{C})}{ds}|_{h=h_{B},s=s_{B}} = 0\;.
\end{align}
Through which, critical temperature and critical classical field value can be obtained. Here, we note that, to ensure two degenerate vacua occur the following constrains also should be satisfied (a positivitive determination of the finite temperature Hessian matrix): $M_1P_1-N_1^2>0,M_1>0$,where
\begin{align}
&\frac{d^2V_{T}(h,s, T_{C})}{dh^2}|_{h=h_{B},s=s_{B}} \equiv M_1 \;, \frac{d^2V_{T}(h,s, T_{C})}{dhds}|_{h=h_{B},s=s_{B}} \equiv N_1\;,\nonumber\\
&\frac{d^2V_{T}(h,s, T_{C})}{ds^2}|_{h=h_{B},s=s_{B}} \equiv P_1\;.
\end{align}
%%%%%%%%%%%%%%%%%%%%%%%%%%%%%%%%%%%%%%%%%%%%%%%%%%%
We first select parameter points met $v_c/T_c>1$ at the critical temperature $T_c$ by using the above methodology, and then study if it is possible to have bubble nucleation, and if the phase transition can complete with {\tt CosmoTransitions}~\cite{Wainwright:2011kj}. In Fig~\ref{onestep}, we show that how the one-step phase transition process works ( $O(0,0)\to B(h,s)$). As the Universe cools down, a second minimum $B(h,s)$ develops, which indicate the break of the $\mathbb{Z} _3$ symmetry and EW symmetry, and the vacuum eventually becomes the present vacuum.

We choose $m_\chi$, $m_2$ ,$v_s$ and $\sin\theta$ as free parameters, and study the phase transition dynamics with these free parameter falls into the following ranges: $m_\chi \in [25,1000]~\rm{GeV}$, $m_2 \in [200,1000]~\rm{GeV}$, $v_s \in [0,500]~\rm{GeV}$ and $|\sin\theta| \leq 0.37$ considering the mixing angle is constrained by the current measurements of the Higgs couplings at the LHC searches~\cite{Ilnicka:2018def}.
For this study, the potential should be bounded from below with,
\begin{equation}
  \lambda_{H} > 0, \quad \lambda_{S} > 0, \quad \lambda_{SH} + 2 \sqrt{\lambda_{H} \lambda_{S}} > 0.
\end{equation}
The unitarity constraints are,
\begin{align}
&  \abs{\lambda_{H}} \leqslant 4 \pi,  \quad\abs{\lambda_{S}} \leqslant 4 \pi, \quad  \abs{\lambda_{SH}} \leqslant 8 \pi,
  \\
&  \lvert 3 \lambda_{H} + 2 \lambda_{S}
   \pm \sqrt{9 \lambda_{H}^2 - 12 \lambda_{H} \lambda_{S} + 4 \lambda_{S}^2 + 2 \lambda_{SH}^{2}} \rvert \leqslant 8 \pi,
\end{align}
%Perturbativity requires~\cite{Lerner:2009xg}: $\abs{\lambda_{H}} \leqslant \frac{2}{3} \pi$, $\abs{\lambda_{S}} \leqslant \pi$ and $\abs{\lambda_{SH}} \leqslant 4 \pi$.
%
%
%
%\begin{figure}[!htp]
%\begin{center}
%\includegraphics[width=0.3\textwidth]{mchi_mh2_theta_vcTc_0111.png}
%\includegraphics[width=0.3\textwidth]{mchi_mh2_vs_vcTc_0111.png}
%\includegraphics[width=0.3\textwidth]{mchi_theta_vcTc_0111.png}
%\caption{Left: the relation between $\theta$ and $m_{\chi}$, $m_{h_2}$ parameters. Middle: we replace $v_s$ as color lable in the same $m_{\chi}-m_{h_2}$ plane. Right: Using these characteristic phase transition strength $v_c/T_c$ as color label in the $m_{\chi}-\theta$(left) plane.}\label{fig:vcTcpTparameter}
%\end{center}
%\end{figure}

%In figure~\ref{fig:vcTcpTparameter}, these points satisfy the SFOEWPT condition $v_c/T_c>1$, with preliminary analysis. the points of Negative $\theta$ only occupy in the $m_{h_2}<400$ GeV, and have large mass $m_{\chi}$. Meanwhile, in most points, it's $m_{\chi}$ heavier than it's $m_{h_2}$. Furthermore, $v_c/T_c$ increase with the mixing angle $\theta$, and most stronger $v_c/T_c$ take place in the $m_\chi>500$ GeV.

\begin{figure}[!htp]
\begin{center}
\includegraphics[width=0.3\textwidth]{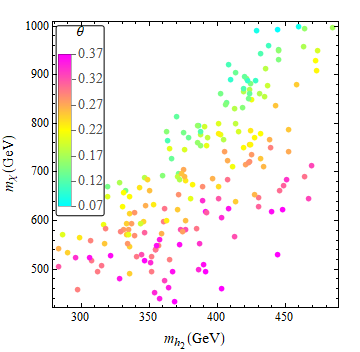}
\includegraphics[width=0.3\textwidth]{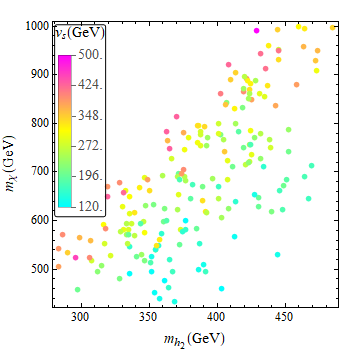}
\includegraphics[width=0.3\textwidth]{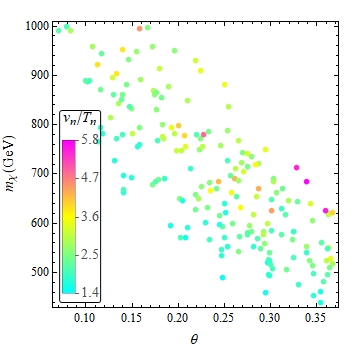}
\caption{Left: the relation among $\theta$, $m_\chi$, and $m_{h_2}$ for EWPT points; Middle: the relation among $v_s$, $m_\chi$, and $m_{h_2}$ for EWPT points; Right: the phase transition strength $v_n/T_n$ as a function of $m_{\chi}$ and $\theta$.}\label{fig:pTparameter}
\end{center}
\end{figure}

In Fig.~\ref{fig:pTparameter}, we show the Electroweak phase transition points. The left (middle) panel indicates that a higher magnitude of $m_\chi$ and $m_{h_2}$ is accompanied with a small mixing angle $\theta$ (a higher magnitude of $v_s$) for the EWPT points. The right panel depicts that a stronger phase transition can be obtained with a large pseudo-scalar mass $m_\chi$ and a large mixing angle $\theta$.

\subsection{BNPC and SFOEWPT}

In this section, we estimate the strongly first-order phase transition condition through the estimation of the BNPC~\cite{Patel:2011th}.
We first calculate the Electroweak sphaleron energy $E_{\rm sph}(T)$ at the phase transition temperature, and then check the relation between the phase transition strength $v(T)/T$
and the following quantity as suggested in Refs.~\cite{Zhou:2020xqi,Zhou:2019uzq},
\begin{align}\label{eq:PTsph1}
PT_{sph}\equiv\frac{E_{\rm sph}(T)}{T} - 7 \ln \frac{v(T)}{T} + \ln \frac{T}{100 \rm{GeV}}\;.
\end{align}
With the quantity $PT_{sph}$ obtained above, we check if the BNPC can met as required by the successful baryon asymmetry generation within the Electroweak baryogengesis~\cite{Patel:2011th,Morrissey:2012db} by the following condition~\cite{Gan:2017mcv}:
\begin{align}\label{eq:PTsph2}
PT_{sph}> (35.9-42.8)\;.
\end{align}
The numerical range here mostly come from the uncertainty of the fluctuation determinant $\kappa = (10^{-4} - 10^{-1})$~\cite{Dine:1991ck}, which is estimated to be comparable with the uncertainty in the lattice simulation of the sphaleron rate at the Standard Model Electroweak cross-over~\cite{DOnofrio:2014rug,Gan:2017mcv}.

\begin{figure}[!htp]
\begin{center}
\includegraphics[width=0.5\textwidth]{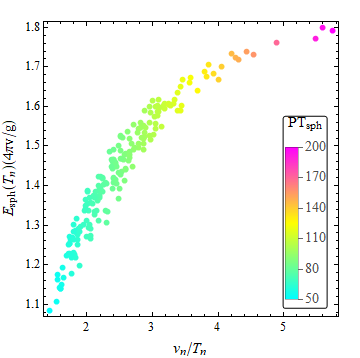}
\caption{The $PT_{sph}$ as a function of the Electroweak sphaleron energy at nucleation temperature and the phase transition strength $v_n/T_n$.  }\label{fig:bnpc}
\end{center}
\end{figure}

In figure~\ref{fig:bnpc}, it is clearly that all the phase transition points satisfy the BNPC, the sphaleron energy $E_{sph}(T_n)$ and $PT_{sph}$ increase with the phase transition strength $v_n/T_n$ increases, and washout of the baryon asymmetry can be
avoided. For uncertainty from Electroweak phase transition duration and bubble nucleation in the evaluation of BNPC, we refer to Ref.~\cite{Patel:2011th}.

\section{Gravitational Waves}

Since, we have the Electroweak symmetry breaking and $\mathbb{Z}_3$ breaking simultaneously, we expect two source of the gravitational radiation at the early Universe.
In this section, we study the stochastic gravitational wave from the strongly first-order Electroweak phase transition and the domain wall decay at latter time.

\subsection{GW from EWPT}

As a crucial parameter for the gravitational wave, $\alpha$ (which is the energy budget of SFOEWPT normalized by the radiative energy) is defined as
\begin{eqnarray}
\alpha=\frac{\Delta\rho}{\rho_R}\;.
\end{eqnarray}
Here, $\rho_R=\pi^2g_\star T_\star^4/30$ is radiation energy of the bath or the plasma background, and $\Delta \rho$ is the latent heat from the phase transition to the energy density of the radiation bath or the plasma background. We take $T_\star\approx T_n$. There is another parameter $\beta$ which characterizes the inverse time duration of the SFOEWPT. Then the GW spectrum peak frequency is defined as
\begin{eqnarray}
\frac{\beta}{H_n}=T\frac{d (S_3(T)/T)}{d T}|_{T=T_n}\; .
\end{eqnarray}
Both the two parameter can be obtained after the solution of the bounce. The action of
the bounce configuration of the field that connects the Electroweak broken vacuum (true vacuum, $\mathbb{Z}_{3}$ broken) and the Electroweak preserving vacuum (false vacuum, $\mathbb{Z}_{3}$ preserving),
\begin{eqnarray}
S_3(T)=\int 4\pi r^2d r\bigg[\frac{1}{2}\big(\frac{d \phi_b}{dr}\big)^2+V(\phi_b,T)\bigg]\;,
\end{eqnarray}
through solving the equation of motion for $\phi_b$ (it is subspace of $h$ and $s$ for this study),
with the boundary conditions of
\begin{eqnarray}
\lim_{r\rightarrow \infty}\phi_b =0\;, \quad \quad \frac{d\phi_b}{d r}|_{r=0}=0\;.
\end{eqnarray}
The phase transition completes at the nucleation temperature when the thermal tunneling probability for bubble nucleation per horizon volume and per horizon time is of order unity~\cite{Affleck:1980ac,Linde:1981zj,Linde:1980tt}:
\begin{eqnarray}\label{eq:bn}
\Gamma\approx A(T)e^{-S_3/T}\sim 1\;.
\end{eqnarray}

\begin{figure}[!htp]
\begin{center}
\includegraphics[width=0.4\textwidth]{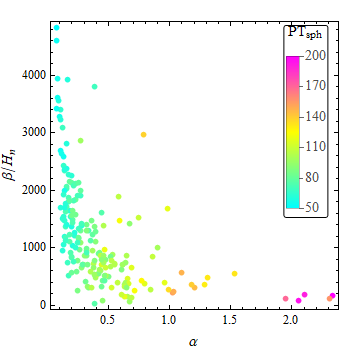}
\includegraphics[width=0.4\textwidth]{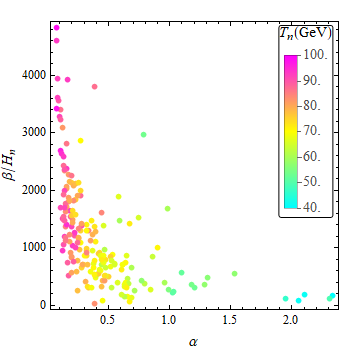}
\caption{Left:  We plot the relations between $\alpha$ and $\beta/H_n$ with the $PT_{sph}$ as color-code; Right: we show GW parameters $\beta/H_n$ and $\alpha$ relation, with nucleation temperature $T_n$ as color-code. }\label{GW:parameter}
\end{center}
\end{figure}

Before going to the study of gravitational wave, we first present the relation between the quantity of $PT_{sph}$ and the two crucial parameters for gravitational wave ($\alpha,\beta/H_n$) in
the left panel of the Fig.~\ref{GW:parameter}. With a smaller $\beta/H_n$ (i.e., a long phase transition duration time) and a larger $\alpha$ (a larger phase transition strength), we obtain a larger $PT_{sph}$. Therefore, the BNPC can be satisfied much better. In the right panel of the Fig.~\ref{GW:parameter}, we present the relation among the nucleation temperature $T_n$ and the gravitational wave ($\alpha,\beta/H_n$). Which depicts that a large $\alpha$ along with a small $\beta/H_n$ can be obtained for a small $T_n$, one can expect a detectable gravitational wave there~\cite{Alves:2018jsw}. Indeed, the two plots also tell that a lower bubble nucleation temperature $T_n$ leads to a larger $PT_{sph}$.

The GWs from the EWPT mainly come from sound waves and MHD turbulence, with
the total energy being given by~\cite{Caprini:2015zlo}
\begin{equation}
\Omega_{\rm GW} h^2 (f) \approx \Omega h^2_{\rm sw} (f) +\Omega h^2_{\rm turb}(f)\;.
\end{equation}

%Firstly, the bubble collisions induced source with the peak frequency locates  \cite{Huber:2008hg}
%\begin{equation}
%f_{\textrm{col}}=
%16.5\times 10^{-6} \frac{0.62}{v_b^2-0.1 v_b+1.8}\frac{\beta}{H}
%\frac{T_*}{100} \left(\frac{g_*}{100}\right)^{\frac{1}{6}} {\rm Hz}\;,
%\end{equation}
%with the energy density being
%\begin{equation}
%\Omega h^2_{\textrm{col}}(f)=1.67\times 10^{-5}\left(\frac{\beta}{H}\right)^{-2}
%\frac{0.11 v_b^3}{0.42+v_b^2}
%\left(\frac{\kappa \alpha }{1+\alpha }\right)^2
%\left(\frac{g_*}{100}\right)^{-\frac{1}{3}}
%\frac{3.8 \left(f/f_{\rm col}\right)^{2.8}}{1+2.8 \left(f/f_{\rm col}\right)^{3.8}},
%\end{equation}
%in which,
Here, we consider detonation bubble and take the bubble wall velocity $v_b$ and the efficiency factor $\kappa$ are functions of $\alpha$~\cite{Steinhardt:1981ct}\footnote{We note that to compatible with EWBG, the wall velocity here can be obtained as a function of $\alpha$.~\cite{Bian:2019zpn,Alves:2018oct,Alves:2018jsw,Alves:2019igs} after taking into account Hydrodynamics. },
\begin{equation}\label{eq:bubblespeed}
v_b=\frac{1/ \sqrt{3}+\sqrt{\alpha ^2+2 \alpha /3}}{1+\alpha }, \quad
%\kappa=\frac{1}{1+0.715 \alpha }\left(0.715 \alpha +\frac{4}{27} \sqrt{\frac{3 \alpha }{2}}\right).
\kappa=\frac{\alpha_{\infty}}{\alpha}\left( \frac{\alpha_{\infty}}{0.73+0.083\sqrt{\alpha_{\infty}}+\alpha_{\infty}} \right).
\end{equation}

The peak frequency of the sound wave locates at\cite{Hindmarsh:2013xza,Hindmarsh:2015qta}
\begin{equation}
f_{\rm sw}=1.9 \times 10^{-5} \frac{\beta}{H} \frac{1}{v_b} \frac{T_*}{100}\left({\frac{g_*}{100}}\right)^{\frac{1}{6}} {\rm Hz }\;,
\end{equation}
with the following energy density being given by
\begin{equation}
\Omega h^2_{\rm sw}(f)=2.65 \times 10^{-6}\left(\frac{\beta}{H}\right)^{-1}
\left(\frac{\kappa \alpha }{1+\alpha }\right)^2
\left(\frac{g_*}{100}\right)^{-\frac{1}{3}}
v_b
\left(\frac{f}{f_{\rm sw}}\right)^3 \left(\frac{7}{4+3 \left(f/f_{\rm sw}\right)^2}\right)^{7/2}.
\end{equation}
Here, the $\kappa$ describes the fraction of the latent heat transferred into the kinetic energy of plasma, we obtain the value by consider the the hydrodynamic analysis~\cite{Espinosa:2010hh}.
The MHD turbulence in the plasma is the second important source of GW signals from phase transition, the peak frequency locates at \cite{Caprini:2009yp}
\begin{equation}f_{\rm turb}=2.7  \times 10^{-5}
\frac{\beta}{H} \frac{1}{v_b} \frac{T_*}{100}\left({\frac{g_*}{100}}\right)^{\frac{1}{6}} {\rm Hz }\;,
\end{equation}
and the energy density is
\begin{equation}
\Omega h^2_{\rm turb}(f)=3.35 \times 10^{-4}\left(\frac{\beta}{H}\right)^{-1}
\left(\frac{\epsilon \kappa \alpha }{1+\alpha }\right)^{\frac{3}{2}}
\left(\frac{g_*}{100}\right)^{-\frac{1}{3}}
v_b
\frac{\left(f/f_{\rm turb}\right)^3\left(1+f/f_{\rm turb}\right)^{-\frac{11}{3}}}{\left[1+8\pi f a_0/(a_* H_*)\right]}\;,
\end{equation}
where the efficiency factor $\epsilon \approx 0.05$,  and the precent Hubble parameter
\begin{equation}
	h_{\ast} = \bigl( 1.65 \times 10^{-5} Hz \bigr) \left( \frac{T_{*}}{100 \rm{GeV}} \right) \left( \frac{g_{\ast}}{100} \right)^{1/6}\;.
\end{equation}
%The detectability of the GWs is quantified by the signal-to-noise ratio (SNR), whose definition is given in
%Ref.~\cite{Caprini:2015zlo}:
%%%%%%%%%%%%%%%%%%%%%%%%%%%%%%%%%%%%%%%%%%%%%%%%%%%%%%%%%%%%%%%%%%%%
%\begin{eqnarray}
%  \text{SNR} = \sqrt{\delta \times \mathcal{T} \int_{f_{\text{min}}}^{f_{\text{max}}} df
%    \left[
%      \frac{h^2 \Omega_{\text{GW}}(f)}{h^2 \Omega_{\text{exp}}(f)}
%  \right]^2} .
%\end{eqnarray}
%Here $h^2 \Omega_{\text{exp}}(f)$ is the experimental sensitivity.
%$\mathcal{T}$ is the mission duration in years for each experiment, assumed to be $5$ here.
%The factor $\delta$ comes from the number of independent channels for cross-correlated
%detectors, which equals $2$ for BBO as well as UDECIGO and $1$ for the others~\cite{Thrane:2013oya}.

\subsection{GW from domain wall decay}

To get domain wall solution formed after the phase transition~\cite{Vilenkin:2000jqa} , we first introduce the phase of the singlet as $S=v_s e^{i\phi}$, and get
the potential of $\phi$ as:
\begin{align}
  V = \frac{\mu_{H}^{2}}{2} v^2 + \frac{\lambda_{H}}{4} v^{4}
  + \frac{\mu_{S}^{2}}{2} v_s^{2} + \frac{\lambda_{S}}{4} v_s^{4} + \frac{\lambda_{SH}}{4} v_s^{2} v^{2} + \frac{\mu_3}{2\sqrt{2}} v_s^3 \cos(3\phi) .
\end{align}
With $\eta^2=v_s^2/2$, the kinetic term of $\phi$ can be obtained as,
\begin{eqnarray}
\mathcal{L}_{\text {kinetic }}(\phi)=\eta^{2}\left(\partial_{\mu} \phi\right)\left(\partial^{\mu} \phi\right)\;.
\end{eqnarray}
The field equation,
\begin{eqnarray}
\partial_{\mu} \frac{\partial \mathcal{L}_{\text {kinetic }}}{\partial_{\mu}(\partial \phi)}+\frac{\partial V}{\partial \phi}=0\;,
\end{eqnarray}
yields
\begin{eqnarray}
\frac{\mathrm{d}^{2} \phi}{\mathrm{d} z^{2}}-\frac{1}{3 B^{2}} \sin (3 \phi)=0\;,
\end{eqnarray}
with
\begin{eqnarray}
\frac{1}{B^{2}}=-\frac{9}{4}\mu_3 v_s^2\;, \phi = \frac{4}{3} \arctan(e^{\frac{z}{B}})\;.
\end{eqnarray}
From which, we can consider a planar domain wall orthogonal to the z-axis~\cite{Hattori:2015xla}, i.e., $\phi(z)$.
The domain wall tension is estimated as,
\begin{eqnarray}
\sigma=\int d z \rho_{\text {wall }}(z)=\int\bigg(\bigg|\frac{dS}{dz}\bigg|^2+V\bigg(\frac{S(z)}{\sqrt{2}},\frac{v}{\sqrt{2}}\bigg)-V\bigg(\frac{v_s}{\sqrt{2}},\frac{v}{\sqrt{2}}\bigg)\bigg) d z\;.
\end{eqnarray}
The same as the previous study of GWs at EWPT, we assume the gravitational radiation produced in the radiation dominated era.  After the formation of the domain wall after the EWPT, one have the domain wall decay. With the peak frequency is given by the Hubble parameter at the decay time~\cite{Hiramatsu:2013qaa}:
\begin{eqnarray}
f^{dw}\left(t_{0}\right)_{\mathrm{peak}}=\frac{a\left(t_{\mathrm{dec}}\right)}{a\left(t_{0}\right)} H\left(t_{\mathrm{dec}}\right)
\simeq 3.99 \times 10^{-9} \mathrm{Hz} \mathcal{A}^{-1 / 2}\left(\frac{1 \mathrm{TeV}^{3}}{\sigma_{\mathrm{wall}}}\right)^{1 / 2}\left(\frac{\Delta V}{1 \mathrm{MeV}^{4}}\right)^{1 / 2}\;,\label{eq:gwfp}
\end{eqnarray}
and peak amplitude of the gravitational waves at the present time $t_0$ is estimated as~\cite{Hiramatsu:2013qaa,Kadota:2015dza}
\begin{eqnarray}
\Omega^{dw}_{\mathrm{GW}} h^{2}\left(t_{0}\right)_{\mathrm{peak}} \simeq 5.20 \times 10^{-20} \times \tilde{\epsilon}_{\mathrm{gw}} \mathcal{A}^{4}\left(\frac{10.75}{g_{*}}\right)^{1 / 3}\left(\frac{\sigma_{\mathrm{wall}}}{1 \mathrm{TeV}^{3}}\right)^{4}\left(\frac{1 \mathrm{MeV}^{4}}{\Delta V}\right)^{2}.\label{eq:gwdw}
\end{eqnarray}
Requiring the domain wall decay before they overclose Universe yields,
\begin{eqnarray}
\sigma_{wall}< 2.93\times 10^4 \mathrm{TeV}^3 \mathcal{A}^{-1} (\frac{0.1 sec}{t_{dec}})\;.
\end{eqnarray}
The bias term $\Delta V$ in Eq.~(\ref{eq:gwfp},\ref{eq:gwdw}) here is introduced to explicitly break the $\mathbb{Z}_3$ symmetry, which determines the decay time of the domain wall,
\begin{equation}
t_{dec}\approx \mathcal{A}\sigma_{wall}/(\Delta V)\;.
\end{equation}
Requiring the domain wall decay before the BBN with $t_{dec}\leq 0.01 sec$~\cite{Kawasaki:2004yh,Kawasaki:2004qu}, one has a lower limit on the magnitude of the bias term:
\begin{eqnarray}
 \Delta V \gtrsim 6.6 \times 10^{-2}\mathrm{MeV}^{4} \mathcal{A} \left(\frac{\sigma_{wall}}{1 \mathrm{TeV}^{3}}\right)\;.
\end{eqnarray}
We note that the magnitude of the bias term should be much less than that of the potential around the core of domain walls ($\Delta V \ll V$ ) such that the discrete $\mathbb{Z}_3$-symmetry holds approximately and not affect the phase transition dynamics. In this study, we take the area parameter $\mathcal{A}=1.2$ for $\mathbb{Z}_3$ symmetry as in Ref~\cite{Kadota:2015dza}, the efficiency parameter $\tilde{\epsilon}_{\mathrm{gw}}=0.7$~\cite{Hiramatsu:2013qaa}, and the degree of freedom at the domain wall decay time $g_{*}= 10.75$ ~\cite{Kadota:2015dza}.
The whole spectrum of the gravitational wave can be obtained after considering the slope of spectrum $\Omega_{GW}^{dw}h^2 \propto f^3$ when $f \textless f_{peak}$, and $\Omega^{dw}_{GW}h^2 \propto f^{-1}$ when $f \geqslant f_{peak}$~as estimated in Ref~\cite{Hiramatsu:2013qaa}.

\begin{figure}[!htp]
\begin{center}
\includegraphics[width=0.45\textwidth]{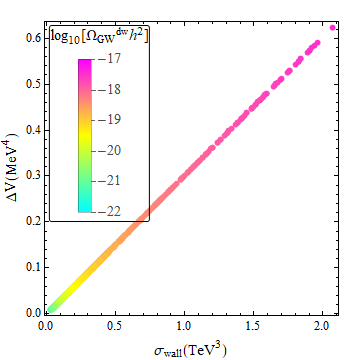}
\caption{We show the relation between the surface mass density $\sigma_{wall}$ and the bias term $\Delta V$, with the $\Omega^{dw}_{GW}h^2$ as the color-code.  }\label{fig:delsig}
\end{center}
\end{figure}

Eq.~\ref{eq:gwfp} and Eq.~\ref{eq:gwdw} indicate that $f^{dw}_{peak}$ is proportional to $\left(\frac{\Delta V}{\sigma_{\mathrm{wall}}}\right)^{1/2}$ and $\Omega_{GW}^{dw}h^2$ is proportional to $\sigma_{\mathrm{wall}}^4$.
In order to evaluate the detectability of the GW from the domain from the SFOEWPT,
we fix $f^{dw}_{peak} \approx 2 \times 10^{-9} Hz$ at the sensitivity frequency of PPTA~\cite{Hobbs:2013aka} to get $\Delta V $ from the the Eq.~\ref{eq:gwfp} with $\sigma_{wall}$ calculated using the SFOEWPT allowed points.
 As we can see in Fig.~\ref{fig:delsig}, with increase of $\sigma_{wall}$ and $\Delta V$, $\Omega_{GW}^{dw}h^2$ can reach to $3.56 \times 10^{-18}$, which is far beyond the sensitivity of the current PTA and the future SKA. Ref.~\cite{Saikawa:2017hiv} shows that a higher magnitude of the gravitational wave spectrum from the domain wall decay requires a large surface mass density of domain walls, which cannot be realized in the SFOEWPT parameter spaces in this model.

\subsection{GW from EWPT with domain wall decay}
\vspace*{0.5cm}
\begin{table}[!h]
\centering
\begin{tabular}{cccccccccccccccccccc}
\hline\hline
  &  $m_{\chi} $~(GeV)  &  $m_{h_2}$~(GeV)  &   $v_s$~(GeV) & $\theta$ & $T_n$~(GeV) & $\beta/H_n$ & $\alpha$\\
%\hline
%$BM_1$  & 463.50 &   537.66  &    166.36    &   0.50 & 55.39 &719.77&0.55\\ [+1mm]
%\hline
$BM_1$  &  625.08 &    361.31  &    184.10 &   0.30 & 50.16 & 219.62 & 1.02\\ [+1mm]
 \hline
$BM_2$  &  814.81 &    370.24  &    243.05 &   0.13 & 69.46 & 152.41 & 0.66\\ [+1mm]
%\hline
%$BM_4$  &  491.47 &    325.43  &    205.39 &  0.42  & 60.94 & 623.13 & 0.45\\ [+1mm]
\hline \hline
\end{tabular}
\def\baselinestretch{1.1}
\caption{Benchmarks in the Fig.~\ref{DWGW-singal}. }
\def\baselinestretch{1.0}
\label{tab:BP}
\end{table}

\begin{figure}[!htp]
\begin{center}
\includegraphics[width=0.6\textwidth]{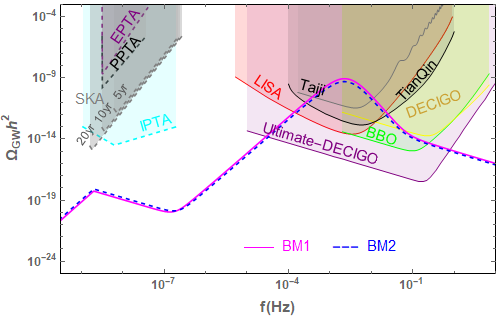}
\caption{Gravitational wave signals from the strong first order electroweak phase transition with domain wall formation and decay. }\label{DWGW-singal}
\end{center}
\end{figure}

We present the gravitational radiation from SFOEWPT and domain wall decay in Fig.~\ref{DWGW-singal} after sum the two contributions. The strength of GW from SFOEWPT is dominant in the higher frequency, and GW strength from the domain wall decay controls the GW spectrum of the low frequency. The gravitational wave signal spectrum with the second higher peak locates around $f_{peak}\sim 10^{-3}-10^{-2}$ Hz can be probed by the projected space-based interferometers, such as: LISA~\cite{Klein:2015hvg}, BBO~\cite{Corbin:2005ny}, DECIGO (Ultimate-DECIGO) \cite{Kudoh:2005as,Musha:2017usi}, TianQin ~\cite{Luo:2015ght} and Taiji \cite{Gong:2014mca} programs. While, the first lower peak from domain wall decay is beyond the sensitivity of EPTA, PPTA, IPTA, and SKA.

\section{Conclusion and discussion}

With the complex singlet scalar preserving $\mathbb{Z}_3$ symmetry, we study the possibility to achieve a one-step strongly first order Electroweak phase transition after considering the baryon number preservation criterion.
After that, we studied the gravitational wave prediction from the strongly first-order Electroweak phase transition with domain wall decay, a two-peak shape is found as expected.
The peak of the predicted gravitational wave signal from the domain wall decay locates around $f_{peak}^{dw}\sim \mathcal{O}(10^{-9})Hz$ with the amplitude of the spectrum cannot be probed by the current sensitivity region of EPTA, PPTA, and IPTA. The peak of the predicted GW spectrum from the phase transition locates at around $f_{peak}^{pt}\sim\mathcal{O}(10^{-3}-10^{-2}) Hz$, with the amplitude within the capability of the projected space-base interferometers, such as: LISA, BBO, DECIGO, UDECIGO, TianQin and Taiji.

\section{Acknowledgements}
This work is supported by the National Natural Science Foundation of China under grant No.11605016 and No.11647307, and the Fundamental Research Fund for the Central Universities of China (No. 2019CDXYWL0029).
We are grateful to Tanmay Vachaspati, Alexander Vilenkin, Ken'ichi Saikawa, Salah Nasri, and Michael J. Ramsey-Musolf for helpful communications and discussions.

\appendix
\section{Vacuum structures and phase transition types}
\label{sec:apa}
\begin{figure}[!htp]
\begin{center}
\includegraphics[width=0.4\textwidth]{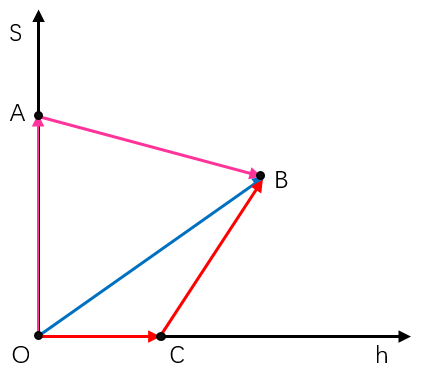}
\includegraphics[width=0.4\textwidth]{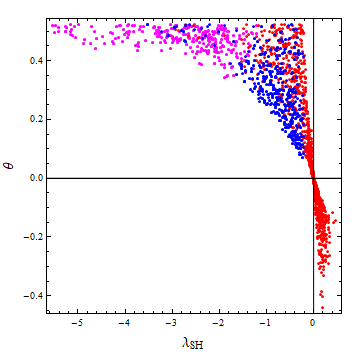}
\caption{Left: The tree level potential vacuum structure of $\mathbb{Z}_3$ model. Right: We show the three types PT points distribution in the $\theta- \lambda_{SH}$ plane.}\label{vacuum}
\end{center}
\end{figure}

As shown in Fig.~\ref{vacuum}, there are totally four possible vacuums locate at $(0,0)$,$(h,0)$,$(0,s)$, and $(h,s)$.
\begin{align}
&O~point~:~ h \to 0, S \to 0\;,\nonumber\\
&A~point~:~ h \to 0, S \to \frac{9 \mu_3^{2}-16 \lambda_S \mu_S^{2}-3 \sqrt{9 \mu_3^{4}-32 \lambda_S \mu_3^{2} \mu_S^{2}}}{16 \lambda_S^{2}}\;, \nonumber\\
&B~point~:~ h \to h_{B}, S \to S_{B}\;, \nonumber\\
&C~point~:~ h \to \sqrt{\frac{-\mu_H^2}{\lambda_H}},S \to 0\; .\nonumber
\end{align}
Therefore, three types phase transition process could happen: One of them is one-step PT ($O \to B$)(Blue), and two-step PT ($O \to A(C) \to B$) (Magenta(Red)). The one-step PT points gather around the $\theta>0$ region. For parameter space with a negative $\theta$, the two-step $O \to C \to B$ could take place. Meanwhile, the two-step PT ($O \to A \to B$) occurs with $\theta\geq 0.35$ that is not favored by the current LHC measurements.
%%%%%%%%%%%%%%%%%%%%%%%%%%%%%%%%%%%%%%%%%%%%%%%%%%%

\section{Electroweak sphaleron}

To compute $E_{sph}(T)$, we obtain the sphaleron solutions following a method suggested in Refs \cite{Manton:1983nd,Klinkhamer:1984di}. Since $U(1)_Y$ contributions are sufficiently small~\cite{Klinkhamer:1990fi,James:1992re}, we employ the spherically symmetric ansatz. Specifically, we consider the configuration of gauge, Higgs and singlet scalar fields are expressed as:
\begin{align}
 A_i(\mu,r,\theta,\phi)
 &=-\frac{i}{g}f(r)\partial_iU(\mu,\theta,\phi)U^{-1}(\mu,\theta,\phi),\\
H(\mu,r,\theta,\phi)
&=\frac{v[T]}{\sqrt{2}}\left[(1-h(r))
	\left(
	\begin{array}{c}
	0 \\
	e^{-i\mu}\cos\mu
	\end{array}
	\right)+h(r)U(\mu,\theta,\phi)
	\left(
	\begin{array}{c}
	0 \\
	1
	\end{array}
	\right)\right],\\
S(\mu,r,\theta,\phi) &= \frac{v_s(T)}{\sqrt{2}} s(r).
\end{align}
where $A_i$ are SU(2) gauge fields, $A_{i}=\frac{1}{2} A_{i}^{a} \tau^{a}$, and $U(\mu,\theta,\phi)$ is defined as
\begin{eqnarray}
U(\mu,\theta,\phi)=
	\left(
	\begin{array}{cc}
	e^{i\mu}(\cos\mu-i\sin\mu\cos\theta) & e^{i\phi}\sin\mu\sin\theta \\
	-e^{-i\phi}\sin\mu\sin\theta & e^{-i\mu}(\cos\mu+i\sin\mu\cos\theta)
	\end{array}
	\right),
\end{eqnarray}
The sphaleron energy in the finite temperature can be written as:
\begin{align}\label{eq:Esph}
	&E_{\rm sph}(T)
	= \frac{4 \pi \Omega[T]}{g} \int_{0}^{\infty} \! d \xi \, \Bigl[
	 4 \biggl( \frac{df}{d\xi} \biggr)^2 s_\mu^2 + \frac{8}{\xi^2} f^2 \bigl( 1 - f \bigr)^2 s_\mu^4 + \frac{\xi^2v[T]^2}{2\Omega[T]^2} \biggl( \frac{dh}{d\xi} \biggr)^2 s_\mu^2\nonumber\\
	&\qquad +\frac{\xi^2v_s[T]^2}{2\Omega[T]^2}\biggl( \frac{ds}{d\xi} \biggr)^2+ s_\mu^2\frac{v[T]^2}{\Omega[T]^2} ( \bigl( 1-f \bigr)^2 h^2  - 2f h (1-f)(1-h)c_\mu^2+f^2(1-h)^2 c_\mu^2)	\nonumber\\ & \qquad
+\frac{\xi^2}{g^2\Omega[T]^4}V_{eff}[\mu,h,s,T]
	\Bigr]
\end{align}
with $\mu \in [0, \pi]$. The configuration at $\mu = \pi/2 $ corresponds to the sphaleron, where $\xi=g \Omega[T] r$, and the $V_{eff}[\mu,h,s,T] = V_{potential}[\mu,h,s,T] - |\Delta[T]|$, and $\Delta[T]$ is the cosmological constant energy density. And the $\Delta[T]$ can be regarded as the minimal value of the potential at temperature $T$. For example, in the $\mathbb{Z}_{3}$ after the temperature cooling at $T=0$ GeV, the constant energy density $\Delta[T]=V_{potential}[\mu,v,v_s,T]$. That is to say, $Min[V_{eff}[\mu,h,s,T]] = 0~\rm{GeV}^4$. The parameter $\Omega[T]$ can take any nonvanishing value of mass dimension one (for example $v[T]$,
$v_S[T]$ or $\sqrt{v[T]^2+v_S[T]^2}$);

From Eq.~(\ref{eq:Esph}), the equations of motion are found to be
\begin{align}
& \frac{d^2f}{d\xi^2}
= \frac{2}{\xi^2}f(1-f)(1-2f)s_\mu^2 +\frac{1}{4}(h^2(f-1)-h(1-h)(1-2f)c_\mu^2 +f(1-h)^2c_\mu^2)\;,\nonumber\\
& \frac{d}{d\xi}\left(\xi^2\frac{dh}{d\xi}\right)
= 2h(1-f)^2   -2f(1-f)(1-2h)c_\mu^2 -2f^2(1-h)c_\mu^2 +\frac{\xi^2}{g^2}\frac{1}{v[T]^2\Omega[T]^2}\frac{\partial V_{\rm eff}}{\partial h}\;,\nonumber\\
&\frac{d}{d\xi}\left(\xi^2\frac{ds}{d\xi}\right)
= \frac{\xi^2}{g^2}\frac{1}{v_S[T]^2\Omega[T]^2}\frac{\partial V_{\rm eff}}{\partial s}\;.
\end{align}
The sphaleron solutions could be obtained with the boundary condition,
\begin{eqnarray}
&&\lim_{\xi\to0} f(\xi) = 0,\quad \lim_{\xi\to0} h(\xi) = 0,\quad \lim_{\xi\to0} s'(\xi) = 0,  \\
&&\lim_{\xi\to\infty} f(\xi) = 1,\quad \lim_{\xi\to\infty} h(\xi) = 1,\quad \lim_{\xi\to\infty} s(\xi) = 1.
\end{eqnarray}

\begin{figure}[!htp]
\begin{center}
\includegraphics[width=0.4\textwidth]{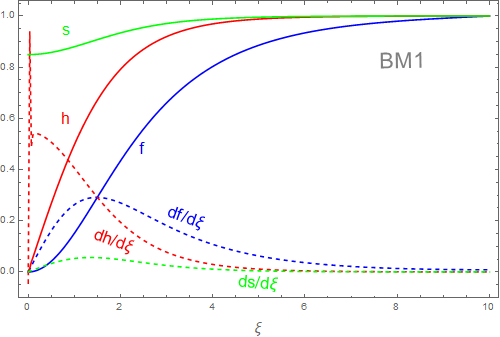}
\includegraphics[width=0.4\textwidth]{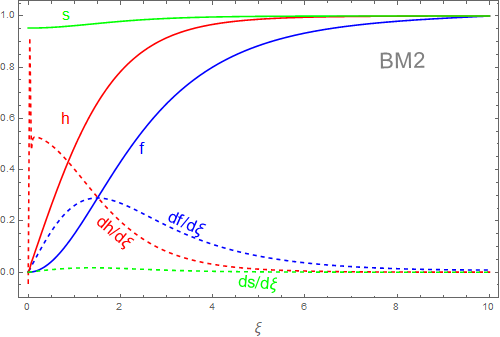}
\caption{Numerically solved Sphaleron($\mu = \pi/2$) profiles of f, h, s and their derivatives as a function of the dimensionless quantity $\xi$ for the two benchmarks in Table~\ref{tab:BP}. }\label{sphaleron}
\end{center}
\end{figure}

\begin{figure*}[!htp]
\centering
\includegraphics[width=0.4\textwidth]{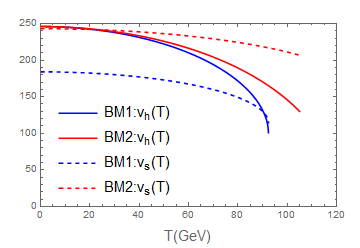}
\includegraphics[width=0.4\textwidth]{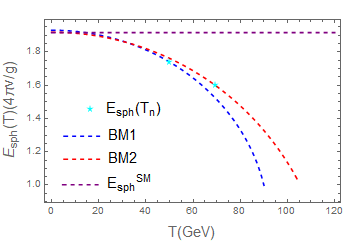}
\caption{Left: Finite temperature VEVs of h(blue) and s(red) versus $T$ for SFOEWPT points.  Right: EW sphaleron energy as a function of temperature. }\label{fig:h6xSMZ3esphT}
\end{figure*}

In Fig.~\ref{sphaleron},  we demonstrate the profile of the Higgs field, $SU(2)$ gauge field, singlet scalar field, and their derivatives behavior, respectively.
In Fig.~\ref{fig:h6xSMZ3esphT}, we illustrate that the sphaleron energy $E_{sph}(T)$ and VEVs of $h(v_h)$ and $s(v_s)$ decrease as the temperature drops for the two benchmark points in Table~\ref{tab:BP}. The sphaleron energy $E_{sph}(T)$ is highly sensitive to the VEV of $h$, as indicated in Ref.~\cite{Patel:2011th}.

%\begin{figure}[!htp]
%\begin{center}
%\includegraphics[width=0.4\textwidth]{Z3_mu_ncs_0921.png}
%\caption{These plots are as a function of Chern-Simons number $N_{cs}$. The energy value is normalized by $4 \pi v[T]/g$. And the brown(dark blue) line shows the $E_{sph}(T_n)$ ($E_{sph}$) behavior with the $N_{cs}$ changing. All these model show that the sphaleron energy in the finite temperature is very closely to the energy in the zero temperature.}\label{vacuum}
%\end{center}
%\end{figure}

\end{document}